\documentclass[twocolumn,english,aps,prl,showpacs,superscriptaddress]{revtex4}

\usepackage{graphicx}
\usepackage{bm}
\usepackage{amsfonts}
\usepackage{amsmath}
\usepackage{color}

\newcommand{\beq}{\begin{equation}}
\newcommand{\eeq}{\end{equation}}
\newcommand{\bea}[1]{\begin{equation}\begin{array}{#1}}
\newcommand{\eea}{\end{array}\end{equation}}
\newcommand{\beqn}{\begin{eqnarray}}
\newcommand{\eeqn}{\end{eqnarray}}

\newcommand{\oper}[1]{\bm{\mathsf{#1}}}

\begin{document}

\title{Reliable Entanglement Detection Under Coarse--Grained Measurements}

\author{D. S. Tasca}
\affiliation{SUPA, School of Physics and Astronomy, University of Glasgow, Glasgow, G12 8QQ, UK}
\affiliation{Instituto de F\'{\i}sica,
Universidade Federal do Rio de Janeiro, Caixa Postal 68528, Rio de
Janeiro, RJ 21941-972, Brazil}
\author{{\L}ukasz  ~Rudnicki}
\email{rudnicki@cft.edu.pl}
\affiliation{Center for Theoretical Physics, Polish Academy of Sciences, Aleja
Lotnik{\'o}w 32/46, 02-668 Warsaw, Poland}
\affiliation{Freiburg Insitute for Advanced Studies, Albert-Ludwigs University of Freiburg, Albertstr. 19, 79104 Freiburg, Germany}

\author{R. M. Gomes}
 \affiliation{Instituto de F\'{\i}sica, Universidade Federal do Rio
              de Janeiro, Caixa Postal 68528, Rio de Janeiro, RJ 21941-972,
              Brazil}
 \affiliation{Instituto de F\'{\i}sica, Universidade Federal de Goi{\'a}s, 74.001-970, Goi\^ania, GO, Brazil}
\author{F.~Toscano}
\affiliation{Instituto de F\'{\i}sica,
Universidade Federal do Rio de Janeiro, Caixa Postal 68528, Rio de
Janeiro, RJ 21941-972, Brazil}

\author{S. P. ~Walborn}
\affiliation{Instituto de F\'{\i}sica,
Universidade Federal do Rio de Janeiro, Caixa Postal 68528, Rio de
Janeiro, RJ 21941-972, Brazil}
\email{swalborn@if.ufrj.br}
\begin{abstract}
We derive reliable entanglement witnesses for coarse--grained measurements
on continuous variable systems. These witnesses never
return a ``false positive" for identification of
entanglement, under any degree of coarse graining.   We show that, even in the case of Gaussian states, entanglement witnesses based on the Shannon entropy can outperform those based on variances.   We apply our results to experimental identification of spatial entanglement of photon pairs.
\end{abstract}

\pacs{03.67.Mn, 03.65.Ud, 42.50.Xa}

\maketitle

\paragraph{Introduction.}
The detection of quantum entanglement is crucial for the implementation of quantum information tasks and technology.  Typically, entanglement is detected through quantum state tomography or the measurement of entanglement witnesses, which involve fewer measurements \cite{duan01,mancini02,horodecki09,guhne09}.  For continuous variable systems, quantum state tomography is difficult due to the large number of measurements required by the infinite dimensional Hilbert space.  Thus, entanglement witnesses are usually more appealing from an experimental point of view.  However, even the experimental measurement of an entanglement witness for high--dimensional systems can be very time demanding.  This is due to the large number of measurements required to reconstruct the probability distributions associated to the continuous variables, for example see \cite{edgar12}.  We note that interesting techniques based on compressed sensing, which reduce the number of measurements, have recently been demonstrated \cite{howland12}.   Nevertheless, it would thus be advantageous to develop entanglement witnesses for coarse--grained measurements.  This would allow for faster accumulation of experimental data with acceptable statistics, as well as for the use of less precise detection schemes.      
\par
The usual entanglement witnesses for continuous variables can fail under general coarse--grained measurements.  In particular, improper application of a CV entanglement witness can result in a ``false positive" in the case of extremely coarse-grained measurements.  In this paper, using recently derived uncertainty relations \cite{rudnicki12a,rudnicki12b}, we develop new sets of entanglement criteria which are acceptable for measurements with any coarse graining.  We illustrate the utility of our approach for spatial variables of entangled photons \cite{walborn10}.  

Let us consider the Hilbert space $\mathcal{H}=\mathcal{H}_{1}\otimes\mathcal{H}_{2}$
of a continuous bipartite state $\hat{\rho}_{12}$. We have coordinate $\left(x_{1},x_{2}\right)$
and momentum $\left(p_{1},p_{2}\right)$ variables associated with
the corresponding Hilbert spaces.  Operators connected with these variables
satisfy usual commutation relations: 
$\left[\oper{x}_{k},\oper{p}_{j}\right]=i\hbar\delta_{kj}$, where $k,j=1,2$.
We will consider entanglement witnesses involving the  global 
{operators}:
\begin{equation}
\oper{x}_{\pm}=\oper{x}_{1}\pm \oper{x}_{2},\qquad \oper{p}_{\pm}=\oper{p}_{1}\pm \oper{p}_{2},\label{global}\end{equation} which obey the commutation relations: { $\left[\oper{x}_{+},\oper{p}_{+}\right]=\left[\oper{x}_{-},\oper{p}_{-}\right]=2 i\hbar$ and $\left[\oper{x}_{+},\oper{p}_{-}\right]=\left[\oper{x}_{-},\oper{p}_{+}\right]=0$}. The marginal probability distributions for the 
bipartite state $\hat{\rho}_{12}$ in the global variables picture (\ref{global}) read:
\begin{subequations}
\begin{eqnarray}
R_{\pm}\left(x_{\pm}\right) & = & \left\langle x_{\pm}\right|\left(\int dx_{\mp}\left\langle x_{\mp}\right|\hat{\rho}_{12}\left|x_{\mp}\right\rangle \right)\left|x_{\pm}\right\rangle ,\label{R+}\\
S_{\pm}\left(p_{\pm}\right) & = & \left\langle p_{\pm}\right|\left(\int dp_{\mp}\left\langle p_{\mp}\right|\hat{\rho}_{12}\left|p_{\mp}\right\rangle \right)\left|p_{\pm}\right\rangle ,\label{S-}\end{eqnarray}\end{subequations}
where $\left|x_{\pm}\right\rangle$ and $\left|p_{\pm}\right\rangle$ are eigenvectors of the operators $\oper{x}_{\pm}$ and $\oper{p}_{\pm}$ respectively.

A { very} useful entanglement witnesses for continuous variables are the Mancini--Giovannetti--Vitali--Tombesi (MGVT) criteria \cite{mancini02,howell04} which state that if $\hat{\rho}_{12}$ is separable then (from now on we set $\hbar=1$):\begin{subequations}
\begin{equation}
\sigma^2[R_{\pm}]\sigma^2[S_{\mp}] \geq 1.
\label{eq:mgvt}
\end{equation}
More sensitive tools are the entropic criteria \cite{walborn09} which in the same situation provide the inequality
\begin{equation}
h[R_{\pm}]+h[S_{\mp}] \geq \ln (2\pi e), 
\label{eq:mgvt-entropic}
\end{equation} \end{subequations}
that is always stronger than (\ref{eq:mgvt}) except for the case of Gaussian states, when both criteria are equivalent.
Here the variance $\sigma^2[f]$ and continuous Shannon entropy $h[f]$ of a probability distribution $f(\cdot)$ are defined in the usual manner:  $\sigma^2[f]=\int dz\, z^{2}f\left(z\right)-\left(\int dz\, zf\left(z\right)\right)^{2}$ and $h[f]=-\int dz\, f\left(z\right)\ln\left[f\left(z\right)\right]$.  Since all separable states must satisfy inequalities (\ref{eq:mgvt}) and (\ref{eq:mgvt-entropic}), their experimental violation is an indication of quantum entanglement.   
\paragraph{Entanglement witnesses under coarse graining.}
In practice, an experiment is performed with finite precision.  In the case of position and momentum of photons or massive particles this is due to the size of the detector.  
The experimental results are then discrete random variables which represent the detection probability for each detection position.  
{ In the case of the measurement of global variables the discrete 
sampling of the continuous distributions comes from the finite precision detectors 
used to measure the position and momentum of each particle.
We can cast this coarse graining if we consider the rectangular function:
\begin{equation}
D_j^\eta\left(z\right)=\begin{cases}
1 & \textrm{ for }z\in\left[\left(j-\frac{1}{2}\right)\eta,\left(j+\frac{1}{2}\right)\eta\right]\\
0 & \textrm{ elsewhere}\end{cases},\label{rect}\end{equation}
{where $\eta$ is the width of the rectangle} and we define $z_{j}=j\eta$ so that $D_j^\eta(z)/\eta \rightarrow \delta(z-z_j)$ as $\eta\rightarrow 0$. Then in global position
and momentum spaces one shall obtain the detection probabilities $\left\{r_{\pm}^{\Delta}\right\}$ and $\left\{s_{\pm}^{\delta}\right\}$ with values
\cite{bialynicki84,bialynicki06,bialynicki11}: 
\begin{equation}
\left\{r_{\pm}^{\Delta}\right\}_k=\!\!\int\limits_{-\infty}^{\infty}\!\!\!dz\, D_k^\Delta(z) R_{\pm}(z),\quad \left\{s_{\pm}^{\delta}\right\}_l=\!\!\int\limits_{-\infty}^{\infty}\!\!\!dz\, D_l^\delta(z) S_{\pm}(z).\label{pr0}\end{equation}
We stress that the widths $\Delta$ and $\delta$ are proportional to the widths of the detectors used to measure position and momentum for each particle respectively.}
The variances of these discrete probability distributions  can be obtained with \cite{rudnicki12a}:
\begin{equation}
\sigma_{r_\pm^{\Delta}}^{2} = \sum\limits_k \left\{r_{\pm}^{\Delta}\right\}_k \left(x_k^\pm\right)^2 - \left ( \sum\limits_k \left\{r_{\pm}^{\Delta}\right\}_k x_k^\pm\right )^2,
\label{eq:varr}
\end{equation}
and corresponding definitions for momentum variances $\sigma_{s_\pm^{\delta}}^{2}$. { The central points $x_k^\pm=k\Delta$ are representative of the
coarse graining global positions measurements. }
Discrete variances are good approximations to the variances of the continuous distributions \eqref{R+} and \eqref{S-} if the
{ widths $\Delta$ ($\delta$) are sufficiently small \cite{rudnicki12a}.}
As $\Delta$ and $\delta$ get large, the variances $\sigma_{r_\pm^{\Delta}}^{2}$ and $\sigma_{s_\pm^{\delta}}^{2}$ tend to zero, since most of the continuous distributions $R_\pm$ and $S_\pm$ 
 will eventually be localized in a single bin.  
However, the experimentalist is limited by the experimental precision of his/her detectors, which are on the order of $\Delta$ and $\delta$.   Thus, the variances calculated according to Eq. \eqref{eq:varr} are not valid estimators of uncertainty in the general case \cite{rudnicki12a,rudnicki12b}.  When the detectors happen to be too large, this can lead to a false detection of entanglement.  For example, consider a pure Gaussian state, with momentum space wavefunction
\begin{equation}
\Psi(p_1,p_2)= A \exp\left(\frac{-(p_1+p_2)^2}{4 \sigma_+^2} \right)\exp\left(\frac{-(p_1-p_2)^2}{4 \sigma_-^2} \right),
\label{eq:psi}
\end{equation}
where $A$ is a normalization constant.   For $\sigma_+=\sigma_-$, the state is separable.  
Nonetheless, if the size of the bins is about $3\sigma_\pm$, all of the probability essentially falls into a single bin for both position and momentum measurements, and the MGVT criteria \eqref{eq:mgvt} can be falsely violated when the variance is calculated according to Eq. \eqref{eq:varr}.    
\par
\begin{figure}
\begin{center}
\includegraphics[width=8cm]{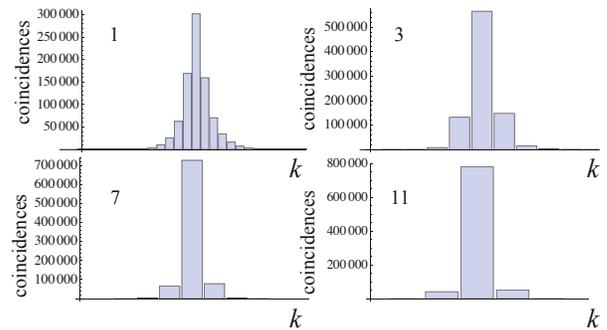}
\end{center}
\caption{\footnotesize  (color online) Binned histogram distributions for experimental results for $x_-$ measurements for different bin sizes $n=1,3,7,11$.}  
\label{fig:xresults}
\end{figure}
\begin{figure}
\begin{center}
\includegraphics[width=8cm]{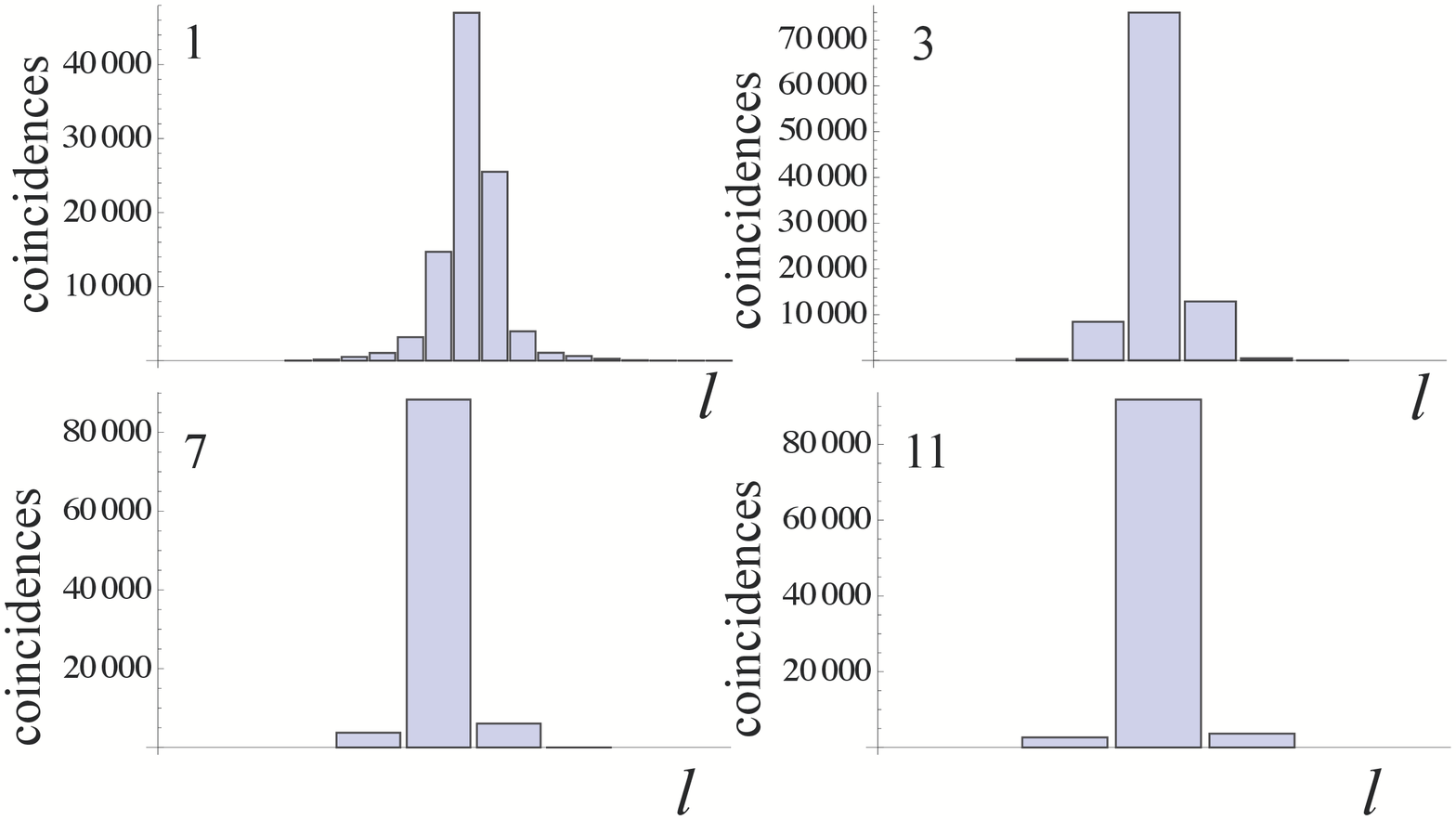}
\end{center}
\caption{\footnotesize  (color online) Binned histogram distributions for experimental results for $p_+$ measurements for different bin sizes $m=1,3,7,11$. }
\label{fig:presults}
\end{figure}

Instead of using the discrete probabilities directly, we can use them to construct approximations to the actual continuous probability distributions for the continuous variables $x_\pm$ and $p_\pm$.   Let us define the distributions:
\begin{subequations}\begin{equation}
R_{\pm}^{\Delta}\left(x_{\pm}\right)=\sum_{k=-\infty}^\infty\left\{r_{\pm}^{\Delta}\right\}_k\frac{D_k^\Delta\left(x_\pm\right)}{\Delta},\label{eq:inferR}\end{equation}
\begin{equation}
S_\pm^{\delta}\left(p_\pm \right)=\sum_{l=-\infty}^\infty\left\{s_{\pm}^{\delta}\right\}_l\frac{D_l^\delta\left(p_\pm\right)}{\delta},
\label{eq:inferS}\end{equation}\end{subequations}
so that $R_{\pm}^{\Delta}$ and $S_\pm^{\delta}$ go to $R_\pm$ and $S_\pm$ in the limit $\Delta,\delta \rightarrow 0$.  The continuous distributions $R_{\pm}^{\Delta}$ and $S_\pm^{\delta}$ are the discretized approximations to $R_\pm$ and $S_\pm$ obtained through coarse--grained measurements.  Figures \ref{fig:xresults} and \ref{fig:presults} show examples of these continuous histogram functions.  

Calculating the variances of the distributions \eqref{eq:inferR} and \eqref{eq:inferS}, one has \cite{rudnicki12a}: $\sigma^{2}\left[R^\Delta_{\pm}\right]  = \sigma_{r_\pm^{\Delta}}^{2} + \Delta^2/12$ and 
$\sigma^{2}\left[S^\delta_{\pm}\right] = \sigma_{s_\pm^{\delta}}^{2}+ \delta^2/12$, 
thus, the discrete variances in general underestimate the inferred variances. 
As $\Delta$ and $\delta$ grow large, these variances are given by $\Delta^2/12$ and $\delta^2/12$, which {represent the variances of the rectangle functions
in Eq.(\ref{rect}) ($\eta=\Delta,\delta$).} 
In a similar fashion, the continuous Shannon entropies of the distributions \eqref{eq:inferR} and \eqref{eq:inferS} are \cite{rudnicki12a}: 
\begin{equation}
h[R_\pm^\Delta] = H\left[\left\{r_{\pm}^{\Delta}\right\}\right] \! + \log \Delta, \quad
h[S_\pm^\delta] = H\left[\left\{s_{\pm}^{\delta}\right\}\right] \! + \log \delta, \label{eq:Hpm}
\end{equation}
where $H\left[\left\{r_{\pm}^{\Delta}\right\}\right]$ and $H\left[\left\{s_{\pm}^{\delta}\right\}\right]$ are Shannon entropies  corresponding to discretizations of the continuous distributions $R_\pm$ and $S_\pm$ \cite{cover}, defined by the well known formula $H[\{q\}]=-\sum_k q_k\ln q_k$. 
{ In the limit of large $\Delta,\delta$ the entropies \eqref{eq:Hpm} are given by the continuous entropies of the rectangle functions, $\log \Delta$ or $\log \delta$. } 
\paragraph{Coarse--grained entanglement criteria.} 
{
The linear relationship between the discrete variances (entropies) and the continuous variances (entropies) allows for a simple 
generalization of the entanglement criteria (\ref{eq:mgvt}) and (\ref{eq:mgvt-entropic}) to the case of coarse grained measurements.  In the supplementary material \cite{sup} we provide a detailed derivation using an improved entropic uncertainty relation \cite{rudnicki12b}.  If state $\hat{\rho}_{12}$ is separable then: }
\begin{subequations}\begin{equation}
\sigma^{2}\left[R_{\pm}^\Delta\right]\sigma^{2}\left[S_{\mp}^\delta\right] -1\geq 0,
\label{eq:genmgvt}
\end{equation}
\begin{equation}
h[R_{\pm}^\Delta]+h[S_{\mp}^\delta] +\ln\left[ \mathcal{C}\left(\delta \Delta \right)\right]\geq0,
\label{unc+-}
\end{equation} 
where \cite{rudnicki12b}
\begin{equation}
\mathcal{C}\left(\gamma\right)=\min\left\{ \frac{1}{2 e\pi};\frac{1}{4\pi}\left[R_{00}\left(\frac{\gamma}{8},1\right)\right]^{2}\right\} .
\label{LR}
\end{equation}
\end{subequations}
Eq. (\ref{eq:genmgvt}) is an entanglement witness based on the variance product, while Eq. (\ref{unc+-}) establishes the entropic entanglement witness, both for the \textit{coarse--grained} probability
distributions. In the limit $\Delta\rightarrow0, \; \delta\rightarrow0$ (\ref{eq:genmgvt}) goes to the MGVT criteria \eqref{eq:mgvt} and (\ref{unc+-}) reproduces the entropic criteria (\ref{eq:mgvt-entropic}). Inequality (\ref{unc+-}) is always stronger than (\ref{eq:genmgvt}) since even for a Gaussian quantum state,  the coarse--grained probability distributions, illustrated in Figures \ref{fig:xresults} and \ref{fig:presults}, for example, are not Gaussians. {Moreover, the improved lower bound in inequality \eqref{unc+-} guarantees that there is no bin size for which it is trivially satisfied \cite{rudnicki12b}. Thus, there is always some entangled state that will violate \eqref{unc+-}.  } 
\paragraph{Experiment.}
\begin{figure}
\begin{center}
\includegraphics[width=8cm]{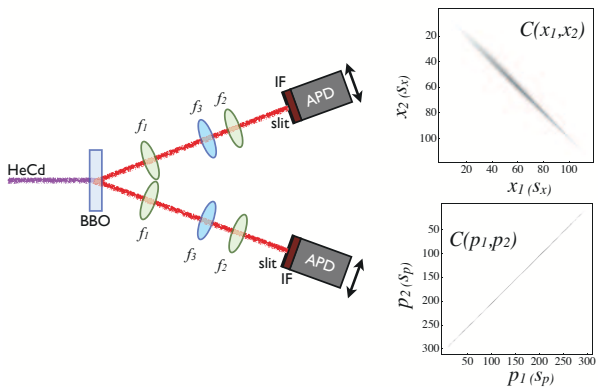}
\end{center}
\caption{\footnotesize  (color online) Experimental setup (left) and joint coincidence distributions (right). Near- and far-field were mapped onto the detection planes by means of switchable lens systems (see text). The distance from the BBO crystal to the detection planes is $500$mm. }
\label{fig:setup}
\end{figure}


We tested the coarse-grained entanglement criteria using spatially entangled photons from spontaneous parametric down-conversion (SPDC), prepared approximately in the state \eqref{eq:psi}. A $5$mm long BBO crystal was pumped with a 325nm cw pump laser in a TEM00 mode.  Down-converted photons at the degenerate wavelength of 650nm were detected  through $10$nm FWHM interference filters using single photon detectors. For a Gaussian pump beam, the transverse spatial structure of the down-converted photon pairs can be approximately described by a Gaussian wavefunction \cite{monken98a, howell04,law04,tasca08, walborn10}, which is factorable in the the $x$ and $y$ cartesian coordinates. Within this configuration, we use narrow slits in the detection system to access one of the transverse dimensions of the two-photon field, whose continuous joint detection probability can be estimated from Eq. \eqref{eq:psi}.

As in Ref. \cite{howell04}, the spatial correlations were measured by using optical lens systems  to map the near-field ($x$) and far-field ($p$) transverse coordinates onto the detection planes.  The $x$ measurements used an imaging system with magnification of $4$, consisting of a telescope with lenses $f_1=50$mm and $f_2=200$mm, while the $p$ measurements used a Fourier transform system with a lens $f_3=250$mm, as illustrated in Figure \ref{fig:setup}. We measured two-dimensional arrays of coincidence counts for the near-field ($x$) and far-field ($p$) variables by scanning the detectors across the vertical direction in the detection planes.  The widths of the slits used in our measurements were $s_x=0.050$mm for $x$ measurements and $s_p=0.020$mm for $p$ measurements. In both cases, the step size used in the scanning was equal to the slit width ($s_x$ or $s_p$). The measured joint probability distributions $C(x_1,x_2)$ and $C(p_1,p_2)$ are shown in the right-hand side of Fig. \ref{fig:setup}.   These were then normalized to obtain probability distributions  $R(x_1,x_2)$ and $S(p_1,p_2)$.  

\par
\begin{figure}
\begin{center}
\includegraphics[width=8cm]{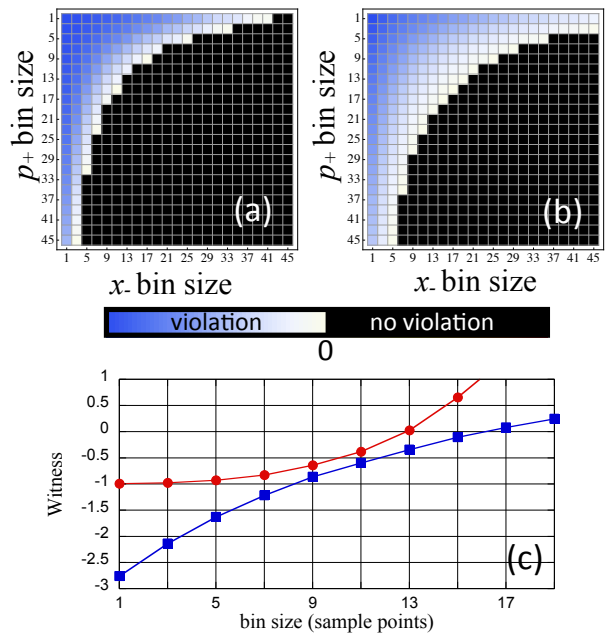}
\end{center}
\caption{\footnotesize  (color online) Evaluation of entanglement witnesses from experimental data as a function of bin size: (a) Variance criteria \eqref{eq:genmgvt} and (b) Entropic criteria \eqref{unc+-}  both as a function of the bin size $n$ and $m$.  Black denotes the region where criteria are not violated. (c) Evaluation of entanglement witnesses for the  same bin size ($n=m$), corresponding to the diagonal in (a) and (b).  The entropic criteria (blue squares) detects entanglement for a larger bin size than the variance criteria (red circles), since the histogram distributions are not gaussian. }
\label{fig:entresults2D}
\end{figure}

{In order to construct different marginal distributions $R_-^{\Delta_n}$ and $S_+^{\delta_m}$ from
different coarse-grained measurements, we calculated these} from the joint distributions in Fig. \ref{fig:setup} by grouping the measurements into different size bins $\Delta_n=n \Delta_1$, $\delta_m=m \delta_1$, where $n,m=1,3,5,7...$.  The smallest bin size is $\Delta_1= 2 s_x (f_1/f_2)=0.0250$mm and $\delta_1=2 s_p(2\pi/f_3 \lambda)=1.546\mathrm{mm}^{-1}$.  Examples of the histogram distributions (before normalization) $R_-^{\Delta_n}$ and $S_+^{\delta_m}$ are shown in Figures \ref{fig:xresults} and \ref{fig:presults}.  Figure \ref{fig:entresults2D} (a) and (b) shows the entanglement witnesses \eqref{eq:genmgvt} and \eqref{unc+-} as a function of the number of bin sizes $n$ and $m$, respectively.  The black region shows the area in which the witnesses do not detect entanglement.  It can be seen that the entropic criteria identifies entanglement for even larger bins than the variance criteria.  This can be seen more distinctly in Figure \ref{fig:entresults2D} (c), which shows these results for the case $n=m$.  As can be seen, even for this approximately Gaussian state the entropic criteria outperform the generalized variance product criteria, due to the coarse graining.         
\par
Error bars, which are smalller than the symbols in Figure \ref{fig:entresults2D} (c), were calculated by propagating the Poissonian counts statistics, as well as the error in the center position of each bin, which we defined to be $\sigma_{\Delta_n}=0.01 \sqrt{2} n f_1/f_2=0.004$mm and $\sigma_{\delta_m}= 0.01 \sqrt{2} (2m\pi/f_3 \lambda)=0.55\mathrm{mm}^{-1}$.  Here $0.01$mm is the minimum step size of the micrometers used to translate the detectors.   
\paragraph{Conclusions.}
Building on previous results for entropic and variance-based uncertainty relations, we have presented entanglement witnesses for continuous variable measurements that are obtained using discretized detection systems.  Our results have been tested in an experiment witnessing spatial entanglement in photon pairs obtained from parametric down-conversion.  Due to the non-Gaussian nature of the binned histogram distributions inferred from discretized measurements, the entropic entanglement witness performs better than the variance-based witness.  {It is important to note that the bin size of the discrete measurements does not alter the entanglement that is in principle available in the system, but rather the experimenter's ability to detect it.} These results should be applicable in other continuous variable quantum systems, {such as entangled atom pairs \cite{perrin07}, time/frequency correlations of photons \cite{shalm12} and superconducting circuits \cite{kirchmair13}.}     
\paragraph{Note added:}  Upon completion of this work, we became aware of similar results obtained for coarse-grained Einstein-Podolsky-Rosen-Steering inequalities \cite{howell12}.   


\begin{acknowledgments}
We thank P.H. Souto Ribeiro for helpful conversations and acknowledge financial support from the Brazilian funding agencies CNPq, CAPES (PROCAD) and FAPERJ.  This work was performed as part of the Brazilian Instituto Nacional de Ci\^{e}ncia e Tecnologia - Informa\c{c}\~{a}o Qu\^{a}ntica (INCT-IQ). Financial support by a grant from the Polish Ministry
of Science and Higher Education for the years 2010\textendash{}2012 and by the European Research Council are gratefully acknowledged. 
\end{acknowledgments}

\end{document}